\documentclass{article}

\pdfoutput=1 
\usepackage{arxiv}
\usepackage[utf8]{inputenc} % allow utf-8 input
\usepackage[T1]{fontenc}    % use 8-bit T1 fonts
\usepackage{hyperref}       % hyperlinks
\usepackage{url}            % simple URL typesetting
\usepackage{booktabs}       % professional-quality tables
\usepackage{amsfonts}       % blackboard math symbols
\usepackage{nicefrac}       % compact symbols for 1/2, etc.
\usepackage{microtype}      % microtypography
\usepackage{lipsum}
\usepackage{amsmath,amssymb}

\usepackage{float}
\usepackage{multirow}
\usepackage{algorithm,algpseudocode}
\usepackage[algo2e]{algorithm2e}
\SetKwRepeat{Do}{do}{while}
\DeclareMathOperator*{\argmin}{arg\,min}

\usepackage{graphicx}
\usepackage{array}
\usepackage{tabu}

\usepackage{xcolor}
\usepackage{multirow}
\usepackage{graphics,graphicx}
\usepackage{multirow}
\usepackage{relsize}
\usepackage{xpatch}
\usepackage{nccmath}
\usepackage{amsmath}

\definecolor{gfet}{HTML}{dcd62f}
\definecolor{gfem}{HTML}{80DC2F}

\def \r{\mathbf{r}}

\title{PAT image reconstruction using augmented sparsity regularization with practicable tuning of regularization weight }

\author{
	Nadaparambil Aravindakshan Rejesh \\
	Imaging Systems Lab\\  Department of Electrical Engineering\\
	Indian Institute of Science (IISc)\\ Bangalore 560012, India\\
	\texttt{rejeshn@iisc.ac.in} \\
		 \And
  Muthuvel Arigovindan
  \thanks{Corresponding author}    \\
  Imaging Systems Lab\\ Department of Electrical Engineering\\
  Indian Institute of Science (IISc)\\ Bangalore 560012, India\\
    \texttt{mvel@iisc.ac.in} \\
}

\begin{document}
\maketitle

\begin{abstract}
Among all tissue imaging modalities,   photo-acoustic tomography (PAT)  has been getting
increasing attention in the recent past due to the fact that it has high contrast,  high penetrability,
and has capability of  retrieving high resolution.  The reconstruction methods used in PAT
plays a crucial role in the applicability of PAT,    and PAT finds  particularly  a wider applicability
if a   model-based regularized reconstruction method is used.  This is because  such methods can yield
high quality images from measurements obtained  using fewer number of transducers.
A crucial factor that determines the quality of reconstruction in such methods is the choice of regularization
weight. Unfortunately,  an appropriately tuned  value of regularization weight varies  significantly
with variation  in  the noise level, as well as, with the variation in the high   resolution contents of the
image, in a way that has not been well understood.  There has been attempts to determine optimum
regularization weight from the measured data  in the context of using elementary and general purpose  regularizations.
In this paper,  we develop a practicable method for semi-automated  tuning of the regularization weight 
in the context of using a modern type of regularization that is specifically designed for PAT
image reconstruction.   As a first step,  we introduce a relative
smoothness constraint with a parameter; this parameter   computationally maps into the actual regularization weight, 
but, its tuning does not vary significantly with variation in  the noise level, and  with the variation in the high 
resolution contents of the image.  Next, we  construct an  algorithm that integrates the task of determining this mapping 
along with obtaining the reconstruction.    Finally we demonstrate experimentally  that we can  run this  algorithm with a 
nominal value of the relative smoothness  parameter---a value  independent of the noise level and the structure of the 
underlying image---to obtain good quality  reconstructions.
\end{abstract}

\section{Introduction}

Photoacoustic tomography (PAT) has received an increasing interest over the past
two decades in soft tissue imaging since it combines the advantages of ultrasound 
imaging and high contrast optical excitation  \cite{wang2016practical,
pramanik2008design,beard2011biomedical, zhou2016tutorial,upputuri2016recent,
li2017single, han2017three}. This hybrid imaging modality has found 
 various applications in biomedicine such as brain imaging in small animals 
 \cite{xia2013small,dean2017advanced}, rheumatoid arthritis research 
 \cite{van2017feasibility,jo2018photoacoustic} and breast imaging in humans 
 \cite{manohar2019current,toi2017visualization,schoustra2019twente}. 
In PAT, some of the absorbed optical energy is converted to heat when the tissue is 
illuminated with nanosecond laser pulses. The resulting thermoelastic expansion inside 
the tissue creates an initial pressure rise;  this propagates as ultrasound waves towards
 the boundary where they are detected by an ultrasound transducer array. 
 These detected signals undergo a reconstruction step to get an estimate of
 2D image of initial pressure-rise,  which represents the spatial distribution of the
 physiological quantity under study. 
 By using this  
 combination of optical absorption and acoustic wave propagation, PAT can retrieve
 higher resolution with higher contrast  from larger depths   compared to purely optical 
 modalities. In most applications of PAT, the high  contrast stems from the  contrast
 in the optical excitability of 
tissue constituents such as haemoglobin. Hence the 
 visualisation of vascular structures by detecting signals from haemoglobin and 
 oxyhaemoglobin becomes possible, which can aid in more robust detection or diagnosis of disorders that
  are characterised by  abnormal proliferation  of blood vessels. 
 
The most common reconstruction methods employed are filtered backprojection (FBP) \cite{finch2004determining, kunyansky2007explicit, xu2002time, xu2005universal,haltmeier2005filtered}  and time reversal \cite{xu2004time,burgholzer2007exact,hristova2008reconstruction,treeby2010k}.
These standard methods suffer from curved line artefacts and blurring, especially when the noise level is high and/or
when the placement of detectors for spatial sampling of the detected photoacoustic (PA) signal is restricted  \cite{rosenthal2013acoustic}. In the past decade, there has been of greater interest in model-based reconstruction methods that  solve  the PAT reconstruction problem by using regularization.
In regularized method, the reconstruction is obtained as a minimizer of  cost that is a   sum of a data-fitting term and a weighted regularization functional or simply a regularization.   The data fitting term measures the  goodness of fit of the candidate image to the measured data via a discrete forward model,  and the regularization  imposes some prior belief on the required solution.  The weight applied on the regularization
is called the regularization weight. Such methods typically yield superior reconstructions even when the number of transducer locations is limited  \cite{buehler2011model,qin2019gradient,yao2011photoacoustic,sanny2018spatially}. Depending on the prior assumptions on the spatial characteristics of the image to be recovered, the regularizers used for these reconstructions have many different forms \cite{buehler2011model,qin2019gradient,yao2011photoacoustic,sanny2018spatially,rejesh2013deconvolution,shaw2013least,huang2013full,arridge2016adjoint,boink2018framework,frikel2018efficient,sandbichler2015novel,haltmeier2016compressed}.  
The traditional choice among these regularization functionals is the Tikhonov regularization functional.
 This functional is  constructed as the  sum of squares of image intensity or image derivatives.  Minimization of this leads to suppression of  noise, and has been  effectively used for PAT image recovery \cite{sanny2018spatially,rejesh2013deconvolution,shaw2013least,haltmeier2015single,dean2012accurate,gutta2018accelerated,prakash2018fractional,bhatt2016exponential}. However minimization of  squared derivatives  leads to over-smoothing;  to alleviate this problem, several improvements were proposed \cite{gutta2018accelerated,prakash2018fractional}. Exponential filtering of singular values in the framework of Tikhonov filtering
was shown to provide better quantitative accuracy  with reduced bias \cite{bhatt2016exponential}.

 Later, sparse regularization approaches gained attention due to their ability to perform well in scenarios where noise is high and/or number transducer locations is low \cite{provost2008application,yao2011photoacoustic,betcke2017acoustic,frikel2018efficient, guo2010compressed,rudin1992nonlinear, arridge2016accelerated,huang2013full, arridge2016adjoint, boink2018framework, han2015sparsity, wang2012investigation, rosenthal2012efficient}.  Early sparsity based methods used wavelet regularization as the prior \cite{provost2008application, guo2010compressed}.
%However, the preservation of structure and detail information in the image is degraded when the wavelet domain regularisers are used for reconstruction from limited data \cite{rosenthal2012efficient}.
Total variation (TV) regularization gained prominence as the demonstration of  its superiority over wavelet regularization
was   facilitated by the developments in convex optimization \cite{yao2011photoacoustic,huang2013full, arridge2016adjoint, boink2018framework, rudin1992nonlinear,  arridge2016accelerated, han2015sparsity, wang2012investigation, zhang2012total}.   The success of TV regularization stems from the fact that  most objects have sparsely distributed derivative values. 
%Wang et al. proposed an adaptive steepest-descent-projection onto convex sets (ASD-POCS) method for the TV-based reconstruction algorithms in three-dimensional PAT \cite{wang2012investigation}.  
Among the TV based methods, first order TV (TV-1) \cite{yao2011photoacoustic,zhang2012total,arridge2016adjoint,huang2013full,arridge2016accelerated,han2015sparsity, wang2012investigation} uses first order derivatives in their formulation, and have been demonstrated   to recover the images with  abrupt edges.  However, when the noise is high and/or number of transducers is low, TV-1 regularization gives reconstructions in the form of piece-wise constants. To alleviate this, second-order TV (TV-2) regularization,
which uses second order derivatives,  was applied in  many imaging inverse problems, and Boink et al.  applied TV-2 for PAT image recovery 
\cite{boink2018framework}.     TV-2  regularization produces reconstruction with reduce artefacts in the presence of  high level of noise and/or high level
of under-sampling,   and it has become one of the most 
widely used regularization for solving imaging inverse
problems.

Further developments includes   total generalised variation (TGV) regularization \cite{bredies2010total}  which helps to combine the best of  TV-1
and TV-2.   It has been applied for PAT   image reconstruction \cite{hammernik2017variational}.   Although TGV combines the best of TV-1 and
TV-2 and hence superior to TV-2 for general image recovery,  a recent study reported by Boink et al \cite{boink2018framework}  indicates that 
TGV's performance is comparable to that of TV-2 in the case of PAT image recovery.  In our previous work \cite{rejesh2020photo},  we  constructed a 
 regularization functional that combines the  intensity and the second-order derivatives to exploit the fact that
  high intensities and high second-order derivatives are 
 jointly  sparse in PAT images.  We proposed a novel optimization method to solve the reconstruction problem using this regularization,  and
 demonstrated that this regularization significantly outperforms TV  regularization. In this paper,  we will call this regularization as
 the augmented sparsity regularization or simply the  augmented regularization.
 
It is important to note that  accurately tuning the regularization weight is crucial for good quality reconstruction in model based method,
and,  in a practical scenario,  the regularization weight  has to be tuned by only using the measured data.
Methods that determine the regularization weight by only using the measured data are sometimes called automated tuning
methods;  because,  without these methods,  the user has to choose the weight by trial-and-error approach where the reconstruction is 
performed for a series of values of the weight 
 with a visual feed back on the quality of the reconstructed image, which is called the manual tuning.
 Clearly, manual tuning   is tedious as the appropriate value of the weight vary drastically
 with variations in the noise level and the structural contents of the underlying image.
 
With our focus on automated tuning of regularization weight, we note that  Prakash et al.  have reported method for automated tuning of regularization 
weight for PAT image recovery using Tikhonov regularization \cite{prakash2018fractional, prakash2014basis,
prakash2019maximum,shaw2013least}.  As sparsity based regularization retrieves better resolution,  our specific focus will be on 
tuning of   regularization weight  for PAT image recovery  in the context of using such regularizations.  With this,
we note that most sparsity based methods proposed for PAT  reconstruction 
 use the  test models themselves for determining the optimal value for regularization.  To be more specific,   these methods simulate 
 measurements from test image
 models,  and adjust the regularization weights such that the  reconstruction obtained from the simulated measurements best matches with the 
 model that generated the measurements.   Although  this way of tuning for regularization weight  is unrealistic since the underlying model image
 is used for tuning,  the goal in these works has been to demonstrate the effectiveness of the regularization.  We will call this way of tuning for 
regularization weight as the {\em oracle tuning}.

There are few dedicated sparsity methods that  deal with the problem of determining the optimal regularization weight 
 from the measured images for elementary inverse  problems such as denoising and deblurring;  they are based on 
 Stein's Unbiased Risk
 Estimator (SURE) principle.   SURE principle   gives a computational expression for means squares error between the
 reconstructed image and the underlying image that generated data without involving the latter in the expression;  SURE
 based method determine the required regularization weight by minimizing this estimate.
 Although this sounds appealing,  the computational expression is very complex; it is complex because it involves derivatives
 of the reconstruction  as a function, where the derivatives are taken with respect to  measured data values. Further,  SURE
 based methods requires knowing the exact values of noise parameter(s). 
 Early SURE methods   were  based on wavelet regularization \cite{surelet, mwsurelet, mwpurelet}.
As the total variation based methods give better results,  the focus moved on to total variation based methods \cite{ mcsure, fengxue}.
The work of Ramani et al. \cite{mcsure}  provides general computational framework for SURE based determination  of
regularization weight for various problems including total variation based denoising.  The work of Xue et al \cite{fengxue}
allows determining the regularization weight for  solving   deblurring  using total variation regularization.   Adopting this
approach for PAT problem is impractical since this method is based on expensive Monte Carlo simulations.

 In this paper,
  we develop a semi-automatic method for determining   the  regularization weight from measured data.
  We do this through the following steps:
 \begin{itemize}
 \item
 We propose a novel smoothness measure,  that is a monotonically decreasing function of the regularization weight, $\lambda$. 
 While appropriate value of $\lambda$ varies drastically with variations in the noise level and structural properties  of
 the underlying image,  this smoothness measure lies within a narrow range if the value of $\lambda$ is chosen appropriately. 
 We propose to use this smoothness measure as the user parameter instead of  using regularization weight $\lambda$ directly.
  \item
 We propose a method that  jointly performs the reconstruction and  determines the  corresponding  value of $\lambda$ for a user defined
 value of the relative smoothness measure. We consider both TV-2 regularization and  a modified   version of   the augmented
 regularization proposed in our recent work \cite{rejesh2020photo}. 
 \item
 We demonstrate experimentally  that we can  run this  algorithm with a fixed nominal value of the relative smoothness parameter 
 to obtain good quality reconstructions  independent of the noise level and the structure of the underlying image.   We compare the 
  structural similarity (SSIM) 
  scores of reconstruction obtained this way to that of reconstructions  obtained with oracle tuning of the regularization weight, and show that
  the SSIM scores are comparable. This means that, in a practical point of view, our work solves the problem of determining the
  required  regularization weight from measured images.
 \end{itemize}
We state that our method is semi-automatic, because, the above-mentioned nominal value for the relative smoothness was
 determined by trial and error method. 
 However, it should be emphasized that  a specific value works for all data sets independent of noise level and the
 underlying image.  Hence our algorithm has practical significance although it is semi-automatic.

\section{Detailed overview of derivative based regularization methods for model based PAT image reconstruction}

The key relation  necessary for the development of regularized reconstruction in PAT is expressing the discrete
image formation model.  To this end,  the first step is expressing the forward model for PAT image reconstruction
in terms of the relation between the laser energy  deposited in the imaging specimen,
$E(\textbf{r},t)$, and pressure field,   $p(\textbf{r},t)$. Here both quantities
are expressed as functions of space and time, with $\r$ representing 2D spatial
location and $t$ representing time.  This relation is given by  \cite{rosenthal2013acoustic}
\begin{equation}
\label{eq:waveeqn}
\Big( \frac{\partial^{2} }{\partial t^{2}} - c^{2}_{0} \; \nabla^{2}  \Big) \; p(\textbf{r},t) 
=  \Gamma \; \frac {\partial }{\partial t} E(\textbf{r},t),
\end{equation}
Here $\Gamma$ is a dimensionless parameter called the Gr$\ddot{u}$neisen coefficient, 
which describes the conversion efficiency of heat to pressure and $c_{0}$ is the speed of
sound in a homogeneous medium.
 By recognizing that the temporal duration of the laser pulse is shorter than the temporal 
 resolution of the ultrasound detectors in most of the photoacoustic imaging applications, 
 the PA source $E(\textbf{r},t)$  may be approximated by  $E(\textbf{r})\delta(t)$, where 
 $E(\textbf{r})$  is the density of deposited energy. Then the solution to the differential 
 equation 
\eqref{eq:waveeqn}  can be written as \cite{rosenthal2013acoustic}
\begin{equation}
\label{eq:contfmodel}
    p(\textbf{r},t) = \cfrac{\Gamma}{4 \pi c_{0} } \;\; \frac{\partial }{\partial t}\; 
    \int\limits_{ |\textbf{r} - \textbf{r}^\prime|= c_{0} t} \; 
    \dfrac{E(\textbf{r}^\prime)}{|\textbf{r} - \textbf{r}^\prime|}\; d\textbf{r}^\prime.
\end{equation}
Here the initial pressure field ${p}_0(\textbf{r})= p(\textbf{r},t=0)$ can be written as
\begin{equation}
  {p}_0(\textbf{r})= \Gamma \;E(\textbf{r}).
\end{equation}
The initial pressure field represents the physical quantity of interest and goal of PAT image
reconstruction task is to recover  ${p}_0(\textbf{r})$ from the samples of $p(\textbf{r},t)$.
The key point that makes this task challenging is that ${p}_0(\textbf{r})$  is required
to be determined for the entire cross-sectional plane, whereas the samples of 
 $p(\textbf{r},t)$ are available only from the points lying in the periphery of the imaging
 specimen where the transducers are located.

To perform the regularized  reconstruction in a computer,  one  needs to  have discrete
representation of the relation given above.  To this end,  the required image itself is represented
as a discrete 2D array,  and the samples of    $p(\textbf{r},t)$  are represented in terms
of numerical approximation of the integral given above.  Let $\{\r_i, i=0,\ldots,L-1\}$ be 
the transducer locations and let $\{t_j, j=0,\ldots,M_t-1\}$ be the sampling  instants.
Further, suppose   ${p}_0(\textbf{r})$ is discretized in a $N_x\times N_y$ cartesian
grid, and suppose ${\bf p}_0$ represents discrete samples of ${p}_0(\textbf{r})$
scanned into vector form with length $N=N_xN_y$.
 Then the  forward model can be represented by 
\begin{equation}
\textbf{m} = \textbf{H} \textbf{p}_0
\end {equation} 
where ${\bf m}$ is the $LM_t\times 1$ vector with $\ell$th element satisfying
 $\{{\bf m}\}_{\ell} = p({\bf r}_i, t_j)$ such that $\ell=M_ti+j$,  and ${\bf H}$ is the matrix representing
 the discrete equivalent of the integral in the equation \eqref{eq:contfmodel}.
Several methods to improve the accuracy of the model by incorporating transducer responses  and
heterogeneities  in the medium have been proposed \cite{huang2013full, wang2011imaging, rosenthal2010fast}.  
Including the measurement noise, the modified imaging 
equation can be written as
\begin{equation}
\textbf{m} = \textbf{H} \textbf{p}_0 + {\pmb \eta},
\end {equation} 
where ${\pmb \eta}$ represents measurement noise, which is Gaussian.

A reconstruction method in PAT aims to recover the initial pressure distribution $\textbf{p}_0$ 
from the noisy transducer measurement data $\textbf{m}$.
 When the number of transducers is less, the PA image  reconstruction problem is ill-posed, 
 and hence constraints
are imposed on the required solution in the form of regularization. 
In this case, the image  reconstruction problem can be treated as an optimization problem 
where the solution is  obtained by minimizing a cost function. The reconstruction problem can be written as
\begin{equation}
\hat{\textbf{p}}_0 = \argmin _{\textbf{x}} \; J(\textbf{x})
\end{equation} where $J(\textbf{x})$ is the cost function and is given by 
\begin{equation}
J(\textbf{x}) = || \textbf{m} - \textbf{H} \textbf{x} ||_{2}^{2} + \lambda \; {R}(\textbf{x}).
\end{equation}
Here ${R}(\textbf{x})$ is the regularization functional and $\|\cdot\|_{2}$ represents the $l_{2}$ norm.  
The regularization parameter $\lambda$ controls the amount of smoothness  in the solution and fidelity to the measured data.
The regularization functional typically should be able to smooth the noise and hence  constructed using derivative terms.  

Tikhonov regularization has been used in limited data cases  
\cite{rosenthal2013acoustic, wang2012investigation} and is given by
\begin{equation}
{R}(\textbf{x}) =\sum_{i}\| \textbf{D}_{o, i}\textbf{x}\|^{2}_{2} = \sum_{r=1}^{N} 
\sum_{i}\big((\textbf{D}_{o,i}\textbf{x})_r\big)^{2}
\end{equation} 
where $(\cdot)_r$ denotes the $r$th component of its vector argument, and $\textbf{D}_{o,i}$ represents the matrix of $i^{th}$ derivative filter of order $o$.  For example, $\textbf{D}_{1,i} , i=1,2$, are the matrix equivalents of filtering by discrete filters that implement the operators $\frac{\partial}{\partial x}$ and  $\frac{\partial }{\partial y}$.   Further, $\textbf{D}_{2,i} , i=1,2,3$, are the matrix equivalents of filtering by discrete filters that implement the operators $\frac{\partial^{2} }{\partial x^{2}}, \frac{\partial^{2} }
{\partial y^{2}}$  and $\sqrt{2}\frac{\partial }{\partial x \partial y}$.
The resulting minimization of the convex quadratic cost function yields a closed form solution given by
\begin{equation}
\hat{\textbf{p}}_{0} = \big[ \textbf{H}^{t}\textbf{H} + \lambda \;\sum_{i}\textbf{D}_{o,i}^{t}\textbf{D}_{o,i}\big] ^{-1} \textbf{H}^{t} \textbf{m}.
\label{Tikh}
\end{equation} 
As emphasized before,  Tikhonov regularization precludes any large derivative values and hence   it tends to smooth edges in the reconstructed image.  Note that we use the matrix formulation for derivatives  for gaining  notational convenience
in describing the minimization method.  The matrices, $\textbf{D}_{o, i}$ represent   filters  that implement 
the derivatives 
$\frac{\partial}{\partial x}$,  $\frac{\partial } {\partial y}$,   $\frac{\partial^{2} }{\partial x^{2}}$,  $ \frac{\partial^{2} }
{\partial y^{2}}$  and $\sqrt{2}\frac{\partial }{\partial x \partial y}$;  these
can be directly applied to the images without building
the matrices.  On the other hand, the data fidelity  term requires building the matrix ${\bf H}$ explicitly.
Total Variation (TV)  regularization 
\cite{tao2009alternating, arridge2016accelerated, wang2008new, huang2013full} 
can be expressed as 
\begin{equation}
{R}_{TV}(\textbf{x})  = \sum_{r=1}^{N}  \sqrt{\sum_{i}\big(\textbf{D}_{o,i}
\textbf{x}\big)_{r}^{2}}
\end{equation} 
  With $o=1$  we get TV-1 and with $o=2$ we get TV-2 regularization.
The recent development,  TGV regularization can be expressed as
\begin{equation}
{R}_{TGV}(\textbf{x})  = 
 \underset{{\bf u}_1, {\bf u}_2} {\operatorname{argmin}}\;\;
{\cal N}_1({\bf x}, {\bf u}_1, {\bf u}_2)  +  {\cal N}_2( {\bf u}_1, {\bf u}_2)
\end{equation}
where
\begin{align}
%\nonumber
{\cal N}_1({\bf x}, {\bf u}_1, {\bf u}_2) =  \sum_{r=1}^{N} 
& \left[  ((\textbf{D}_{1,1}\textbf{x}\big)_{r}- ({\bf u}_1)_r)^2   \right.  \left. +  ((\textbf{D}_{1,2}\textbf{x}\big)_{r}- ({\bf u}_2)_r)^2   \right]^{0.5}
 \end{align}
\begin{align}
%\nonumber
{\cal N}_2( {\bf u}_1, {\bf u}_2) =  \sum_{r=1}^{N} 
& \left[  ((\textbf{D}_{1,1}\textbf{u}_1\big)_{r})^2 +
((\textbf{D}_{1,2}\textbf{u}_2\big)_{r})^2  + 
  \right.  \left. + 0.5((\textbf{D}_{1,2}\textbf{u}_1\big)_{r}+(\textbf{D}_{1,1}\textbf{u}_2\big)_{r})^2  \right]^{0.5}
 \end{align}
with  ${\bf u}_1$ and ${\bf u}_2$ being $N\times 1$ vectors.  Note that the TGV regularization is expressed 
via minimization
in $2N$ dimensional space.

The photoacoustic images have high contrast due to the differential absorption of light in the near-infrared region by chromophores such as hemoglobin. Due to this,  high values of initial pressure, $p_0(\mathbf{r})$, are sparsely distributed.  Further,  regions having high derivative values are also sparsely distributed.  This pattern was also observed for fluorescence images in the work presented in \cite{arigovindan2013high},   where the regularization was constructed by adding an intensity term to second-order derivatives.  The combined point-wise cost went into a logarithmic function and summed over all pixels.  Inspired by the success of this form of regularization,  we proposed a modification
that suits PAT reconstruction problem in \cite{rejesh2020photo}.    We  replaced the log by a fractional power and  expressed regularization as
\begin{equation}
\label{eq:rh1}
{R}_{h}(\textbf{x}, q)  = \sum_{r=1}^{N}   \left(\epsilon_d + \alpha \big(\textbf{x}\big)_{r}^{2} 
+ (1-\alpha)\sum_{i}\big(\textbf{D}_{o,i}\textbf{x}\big)_{r}^{2}\right)^q,
\end{equation} 
where the weight $\alpha \in (0,1)$ controls the relative penalization, and $\epsilon_d$ is a small positive real number.
   The advantage of this modification  is that it allowed
 an optimization strategy that can efficiently handle non-convex cost function.
 %, which will be demonstrated later. 
We chose $q < 0.5$ meaning that the resulting cost functional  is non-convex.  
We also proposed an optimization method for PAT image reconstruction using the above regularization and demonstrated
their effectiveness using numerical experiments.  We call  ${R}_{h}(\textbf{x}, q)$ as the augmented 
sparsity regularization.

\section{Proposed method}

\subsection{Cost  function and reconstruction algorithm}

Our goal is to make the augmented sparsity regularization more practicable by constructing an iterative
method that integrates the task of determining the regularization weight and  obtaining  regularized reconstruction.
This is challenging, and to make this  computationally tractable,  we simplify the augmented sparsity
regularization by making it convex.  Specifically,  we simplify the regularization given in the equation
\eqref{eq:rh1} as given below:
\begin{equation}
\label{eq:rh2}
{R}_{hc}(\textbf{x}, \alpha)  = \sum_{r=1}^{N}   \left(
\alpha \big(\textbf{x}\big)_{r}^{2} + 
(1-\alpha)\sum_{i=1}^3\big(\textbf{D}_{2,i}\textbf{x}\big)_{r}^{2}\right)^{1/2}.
\end{equation} 
Note that, we have also eliminated $\epsilon_d$ required for making the functional differentiable,  since
non-differentiability can be easily handled when the function is convex.
With this,   the required reconstruction is expressed as 
\begin{equation}
\hat{\textbf{x}} = \argmin _{\textbf{x}} \; J_c(\textbf{x}, {\bf H}, {\bf m}, \lambda, \alpha)
\end{equation} 
where $J_c(\textbf{x}, {\bf H}, {\bf m}, \lambda, \alpha)$ is the cost function and is given by 
\begin{equation}
\begin{split}
J_c(\textbf{x}, {\bf H}, {\bf m}, \lambda, \alpha) =   & \frac{1}{n} \| \textbf{m} - \textbf{H} \textbf{x} \|_{2}^{2}  + \lambda \; {R}_{hc}(\textbf{x}, \alpha) + {\cal B}_{[u]}({\bf x}),
 \end{split}
\end{equation}
with  $n$ denoting the length of the measurement vector ${\bf m}$.
Further, 
${\cal B}_{[u]}(\cdot)$ is the cost function that enforces the candidate image pixels ${\bf x}$ to be
in the range $[0,u]$, where $u$ is the user-defined upper bound.  
If any of the pixels of ${\bf x}$   is outside this range,  ${\cal B}_{[u]}(\cdot)$ becomes $\infty$,
and it will have zero value otherwise.  This function imposes exact bound constraint on the required reconstruction.
Note that, in our original formulation given in \cite{rejesh2020photo},  we
only incorporated an approximate bound constraint, and here we have the exact bound constraint. 
This is again due to the fact that the exact bound constraint is easy to implement since the overall cost function
is convex.  The above minimization problem is very similar to the case where TV-2 functional is 
used instead of ${R}_{hc}(\textbf{x}, \alpha)$. However,  in order to develop the    reconstruction
method that integrates the task of determining the regularization weight, we need to write down
a specific algorithm for minimizing $J_c(\cdot)$.  

For minimizing this cost,  we adopt  alternating direction method of multipliers (ADMM) approach \cite{eck}.
Implementation of ADMM requires separating the sub-functionals of $J_c(\cdot)$
with auxiliary variables that are linked to the primary variable ${\bf x}$ in terms of 
linear equations.  To this end,  we define   $${\bf D}_\alpha = \left[
\begin{array}{c}
\sqrt{\alpha}\;{\bf I}  \\  \sqrt{(1-\alpha)}\;{\bf D}_{2,1}  \\ \sqrt{(1-\alpha)}\;{\bf D}_{2,2} \\ \sqrt{(1-\alpha)}\;{\bf D}_{2,3} 
\end{array}
\right].$$
With this,  ${R}_{hc}(\textbf{x}, \alpha)$ given in the equation can be expressed as
\begin{equation}
\label{eq:rh3}
{R}_{hc}(\textbf{x}, \alpha)  = \sum_{r=1}^{N}   \left(
\sum_{j=0}^3 ({\bf D}_\alpha{\bf x})^2_{r+jN}\right)^{1/2}.
\end{equation} 
To further simplify the notation,   we introduce the interlaced slicing operation given by
$$({\bf v})_{[r]} = [({\bf v})_{r}\;  ({\bf v})_{r+N}\;({\bf v})_{r+2N}\;({\bf v})_{r+3N}]^t.$$
Note that  ${\bf v}$ is in $\mathbb{R}^{4N}$ and $({\bf v})_{[r]}$ is in $\mathbb{R}^4$  for
$r=1,\ldots,N$.   With this,  we can express  ${R}_{hc}(\textbf{x}, \alpha)$  as
\begin{equation}
\label{eq:rh4}
{R}_{hc}(\textbf{x}, \alpha)  = \sum_{r=1}^{N} 
\|({\bf D}_\alpha {\bf x})_{[r]} \|_2
\end{equation} 
For notational convenience in developing the ADMM algorithm, we define
following function for a vector ${\bf v} \in \mathbb{R}^{4N}$:
\begin{equation}
\label{eq:Ndef}
{\cal S}({\bf v}) = \sum_{r=1}^{N} 
\|({\bf v})_{[r]} \|_2.
\end{equation}
With this, the minimization problem can be expressed as
\begin{align}
\label{eq:cprob}
\nonumber
({\textbf{x}}^*, {\bf b}^*, {\bf d}^*)   = & \argmin _{\textbf{x}, {\bf d}, {\bf b}} \;  
\frac{1}{n} \| \textbf{m} - \textbf{H} \textbf{x} \|_{2}^{2}  
 +   \lambda {\cal S}({\bf d}) +  {\cal B}_{[u]}({\bf b}),  \; \;   
 s.t.  \;\;   {\bf D}_\alpha{\bf x} = {\bf d}, \;\;   {\bf x} = {\bf b}.
\end{align} 
The solution to the above constrained optimization problem can be obtained through 
the so-called augmented Lagrangian 
which can be written as follows:
\begin{equation}
\label{eq:auglag}
\begin{split}
L({\bf x}, {\bf y}, \hat{\bf y}) =   & \frac{1}{n} \| \textbf{m}- \textbf{H} \textbf{x} \|_{2}^{2}  +  \bar{\cal S}({\bf y})   +   
\frac{\beta}{2}\|\bar{\bf D}_\alpha{\bf x}-{\bf y}\|_2^2 + \hat{y}^t\left( \bar{\bf D}_\alpha{\bf x} -  {\bf y}\right),
\end{split}
\end{equation}
where $\bar{\bf D}_\alpha = 
 \left[
\begin{array}{c}
{\bf I}  \\  {\bf D}_\alpha
\end{array}
\right]$,  and   ${\bf y} = 
 \left[
\begin{array}{c}
{\bf b}  \\  {\bf d}
\end{array}
\right]$, 
and $\hat{\bf y}$ is Lagrange's multiplier vector  with dimension $5N$.   Here, $\beta$
is a penalty parameter,  and
$\bar{\cal S}({\bf y}) = \lambda {\cal S}({\bf d}) + {\cal B}_{[u]}({\bf b})$.
Constrained optimization theory suggests that,  with $\hat{\bf y}$ correctly chosen, say
$\hat{\bf y}=\hat{\bf y}^*$,  minimizing  $L({\bf x}, {\bf y}, \hat{\bf y})$ with respect to
${\bf x}$ and ${\bf y}$ solves the optimization problem given in the equation \eqref{eq:cprob} \cite{Bertsekas99}.
This $\hat{\bf y}^*$ is the maximizer of $q(\hat{\bf y})$, which is the so-called the dual function of 
the constrained optimization problem; this dual function is  given by
\begin{equation}
\label{eq:dualdef}
q(\hat{\bf y}) = \min_{{\bf x},{\bf y}} 
 \frac{1}{n} \| \textbf{m} - \textbf{H} \textbf{x} \|_{2}^{2}  +  \bar{\cal S}({\bf y}) +   
\hat{y}^t\left( \bar{\bf D}_\alpha{\bf x} -  {\bf y}\right).
\end{equation}
The  method of ADMM allows to determine  $\hat{\bf y}^*$ jointly with required solution defined by
$$({\textbf{x}}^*, {\bf y}^*)  = \argmin _{\textbf{x}, {\bf y}}  L({\bf x}, {\bf y}, \hat{\bf y}^*),$$
in terms of simple elementary minimizations without explicitly determining $q(\hat{\bf y})$.
Given a current estimate of the solution, say,  $({\bf x}^{(k)}, {\bf y}^{(k)}, \hat{\bf y}^{(k)})$, 
the next refined estimate in ADMM is obtained as  follows:
\begin{align}
\label{eq:admmx1}
\textbf{x}^{(k+1)}   & = \argmin _{\textbf{x} \in \mathbb{R}^N}  L({\bf x}, {\bf y}^{(k)}, \hat{\bf y}^{(k)}),  \\
\label{eq:admmy1}
\textbf{y}^{(k+1)}   & = \argmin _{\textbf{y} \in \mathbb{R}^{5N}}  L({\bf x}^{(k+1)}, {\bf y}, \hat{\bf y}^{(k)}),  \\
\hat{\bf y}^{(k+1)}  & = \hat{\bf y}^{(k)} + \beta(\bar{\bf D}_\alpha{\bf x}^{(k+1)}-{\bf y}^{(k+1)}). 
\end{align}
The formula that gives the solution for  first minimization problem can be  easily obtained
because $L(\cdot, \cdot, \cdot)$ is quadratic with respect to  ${\bf x}$.    The equation
that represents the minimum is given by
\begin{equation}
\label{eq:xkform}
\left({\bf H}^t{\bf H} + \beta \bar{\bf D}^t_\alpha\bar{\bf D}_\alpha \right){\bf x}^{(k+1)} = {\bf H}^t{\bf m} + \bar{\bf D}^t_\alpha\left(\beta {\bf y}^{(k)} - \hat{\bf y}^{(k)} \right).
\end{equation}
The above equation can be solved in either of two ways.  If the image to be reconstructed
is of moderate size,  inverse of  ${\bf H}^t{\bf H} + \beta \bar{\bf D}^t_\alpha\bar{\bf D}_\alpha$ can be
precomputed outside the ADMM loop,  and the required solution in the ADMM loop
can be computed as ${\bf x}^{(k+1)} =    \left({\bf H}^t{\bf H} + \beta \bar{\bf D}_\alpha^t\bar{\bf D}_\alpha \right)^{-1} 
\left({\bf H}^t{\bf m} + \bar{\bf D}_\alpha^t\left(\beta {\bf y}^{(k)} - \hat{\bf y}^{(k)} \right)\right)$. 
Otherwise,  the equation \eqref{eq:xkform} can be solved for ${\bf x}^{(k+1)}$ using the method of conjugate
gradients.  Next, from the structure of the function $L(\cdot,\cdot, \cdot)$,  we deduce that
the minimization problem given in the equation \eqref{eq:admmy1} can be equivalently
expressed as 
\begin{equation}
\label{eq:admmy2}
\textbf{y}^{(k+1)}    = \argmin _{\textbf{y} \in \mathbb{R}^{5N}}    \;\;  \bar{\cal S}({\bf y}) + \frac{\beta}{2}\|{\bf y}-\bar{\bf y}^{(k+1)}\|_2^2
\end{equation}
where 
$\bar{\bf y}^{(k+1)} = \bar{\bf D}_\alpha{\bf x}^{(k+1)} + \frac{1}{\beta} \hat{\bf y}^{(k)}$.  Note that 
$\bar{\cal S}({\bf y}) = \lambda {\cal S}({\bf d}) + {\cal B}({\bf b})$ and 
${\bf y} = 
 \left[
\begin{array}{c}
{\bf b}  \\  {\bf d}
\end{array}
\right]$.    Let 
${\bf y}^{(k+1)} = 
 \left[
\begin{array}{c}
{\bf b}^{(k+1)}  \\  {\bf d}^{(k+1)}
\end{array}
\right]$
and
$\bar{\bf y}^{(k+1)} = 
 \left[
\begin{array}{c}
\bar{\bf b}^{(k+1)}  \\  \bar{\bf d}^{(k+1)}
\end{array}
\right]$ denote the corresponding partitions.   Then the minimization problem of equation \eqref{eq:admmy2}
can be written as the following sub-problems:
\begin{align}
\label{eq:addmmys1}
\textbf{d}^{(k+1)}    & = \argmin _{\textbf{d} \in \mathbb{R}^{4N}}    \;\;  \lambda {\cal S}({\bf d}) +\frac{\beta}{2} \|{\bf d}-\bar{\bf d}^{(k+1)}\|_2^2. \\
\label{eq:admmys2}
\textbf{b}^{(k+1)}    & = \argmin _{\textbf{b} \in  \mathbb{R}^N}    \;\;  {\cal B}({\bf b}) + \frac{\beta}{2}\|{\bf b}-\bar{\bf b}^{(k+1)}\|_2^2. 
\end{align}
The solution to the second sub-problem is well-known and can be represented by
$\textbf{b}^{(k+1)} = {\cal P}(\bar{\bf b}^{(k+1)})$ where   ${\cal P}(\cdot)$. denotes the clipping of the components
of the vector by the range $[0, u]$.        Next, we note that the first sub-problem is separable with respect to 
the slicing operation $(\cdot)_{[r]}$. In other words, the first subproblem can be written as
\begin{equation}
\label{eq:admmys1n}
(\textbf{d}^{(k+1)})_{[r]}    = \argmin _{\textbf{v}\in \mathbb{R}^4}     \lambda  \|{\bf v}\|_2 + 
\frac{\beta}{2} \left\|{\bf v}-(\bar{\bf d}^{(k+1)})_{[r]}\right\|_2^2.
\end{equation}
This means that we can compute $(\textbf{d}^{(k+1)})_{[r]}$ for $r=1,\ldots,N$ and assemble to get the solution
$\textbf{d}^{(k+1)}$.    The solution to the problem of equation \eqref{eq:admmys1n} is well known and can be
expressed as  \cite{amirbeck} 
$(\textbf{d}^{(k+1)})_{[r]}    =   \textbf{Prox}_{l_2}((\bar{\bf d}^{(k+1)})_{[r]}, \lambda / \beta )$, where 
$\textbf{Prox}_{l_2}({\bf z}, \lambda)$ is given by
\begin{equation}
\label{eq:admmys1ns}
\textbf{Prox}_{l_2}({\bf z}, \lambda)   =   \frac{\max(\|{\bf z}\|_2-\lambda, 0)}{\|{\bf z}\|_2}{\bf z} 
\end{equation}
Considering the choice of $\beta$,  it is sufficient to set it equal to $1$ for all measured datasets, although 
 minor improvement in the convergence speed can be obtained by adjusting the value.

\subsection{Modified algorithm with  relative smoothness constraint }

To introduce the idea of relative  smoothness constraint,   let ${\bf H}_f$ be the given measurement
matrix and let  ${\bf H}$ be the matrix obtained by removing a fraction of rows equal to $\Delta$.
In other words, $n/n_f = 1 - \Delta$ where $n$ and $n_f$ are number of rows in the matrices
${\bf H}$ and ${\bf H}_f$ respectively.
Let ${\bf m}_f$ and ${\bf m}$ be the corresponding measurement vectors.  Let ${\bf x}(\lambda)$
denote the reconstruction obtained from ${\bf m}$  by using  the regularization weight $\lambda$.
In other words,  let
\begin{align}
{\textbf{x}}(\lambda) = & \argmin _{\textbf{x}}   \;\;  
J_c(\textbf{x}, {\bf H}, {\bf m}, \lambda, \alpha).
\end{align}
where 
\begin{align}
\nonumber
J_c(\textbf{x}, {\bf H}, {\bf m}, \lambda, \alpha)    =  &
\frac{1}{n}\| \textbf{m} - \textbf{H} \textbf{x} \|_{2}^{2}   + \lambda \; {R}_{hc}(\textbf{x}, \alpha) + {\cal B}_{[u]}({\bf x}).
\end{align}
Consider
\begin{align}
\nonumber
J_c(\textbf{x}, {\bf H}_f, {\bf m}_f, \lambda, \alpha)    = & 
\frac{1}{n_f} \| \textbf{m}_f - \textbf{H}_f \textbf{x} \|_{2}^{2}  + \lambda \; {R}_{hc}(\textbf{x}, \alpha) + {\cal B}_{[u]}({\bf x}).
\end{align}
Now, our hypothesis is the following:     if   $\lambda$ is chosen such that ${\bf x}(\lambda)$
does not fit noise,  then the data fidelity terms 
$\frac{1}{n}\| \textbf{m} - \textbf{H} {\bf x}(\lambda) \|_{2}^{2}$  and 
$\frac{1}{n_f} \| \textbf{m}_f - \textbf{H}_f{\bf x}(\lambda) \|_{2}^{2}$  will also be nearly
identical because of the normalization with number of measurements.  
However, in practice,   the difference $\left\|\frac{1}{n}\| \textbf{m} - \textbf{H} {\bf x}(\lambda) \|_{2}^{2} - 
\frac{1}{n_f} \| \textbf{m}_f - \textbf{H}_f{\bf x}(\lambda) \|_{2}^{2}\right\|$
will  be monotonically decreasing.  As a result,  the  difference
$|J_c({\bf x}(\lambda), {\bf H}, {\bf m}, \lambda, \alpha) 
- J_c({\bf x}(\lambda), {\bf H}_f, {\bf m}_f, \lambda, \alpha)|$
will  also be monotonically
decreasing w.r.t. $\lambda$.
Based on these observations,  we define the following:
 \begin{equation}
\label{eq:sdef}
{S}(\lambda) =  \frac{|{\cal J}_f({\textbf{x}}(\lambda), \lambda) - 
{\cal J}({\textbf{x}}(\lambda), \lambda)|}
{0.5({\cal J}_f({\textbf{x}}(\lambda), \lambda) +
{\cal J}({\textbf{x}}(\lambda), \lambda))},
\end{equation}
where ${\cal J}_f({\textbf{x}}(\lambda), \lambda) = J_c({\bf x}(\lambda), {\bf H}_f, {\bf m}_f, \lambda, \alpha)$
and 
${\cal J}({\textbf{x}}(\lambda), \lambda) = J_c({\bf x}(\lambda), {\bf H}, {\bf m}, \lambda, \alpha)$.
We propose that the user can specify the required smoothing indrectly as an upper bound 
on  ${S}(\lambda)$, say $\epsilon$, and the required value for $\lambda$ can be determined
as the lowest value of $\lambda$ that satisfy  ${S}(\lambda) \le \epsilon$.   The main advantage
is that  ${S}(\lambda)$ is a measure of adeqaucy  of the magnitude of $\lambda$ for a given
measured data set,  and variation in the reconstructed image quality with respect to changes
in ${S}(\lambda)$  will be much narrower than the variation in the  reconstructed image quality
 with respect to changes in $\lambda$.

\begin{figure}[htp] 
 \caption{Reconstruction   with regularization weight tracking.  $\frac{n}{n_f} = 1 -\Delta$ where  $n$
 and $n_f$  are number of elements in ${\bf m}$ and ${\bf m}_f$ respectively.}
 \centering
 \includegraphics[width=0.5\textwidth]{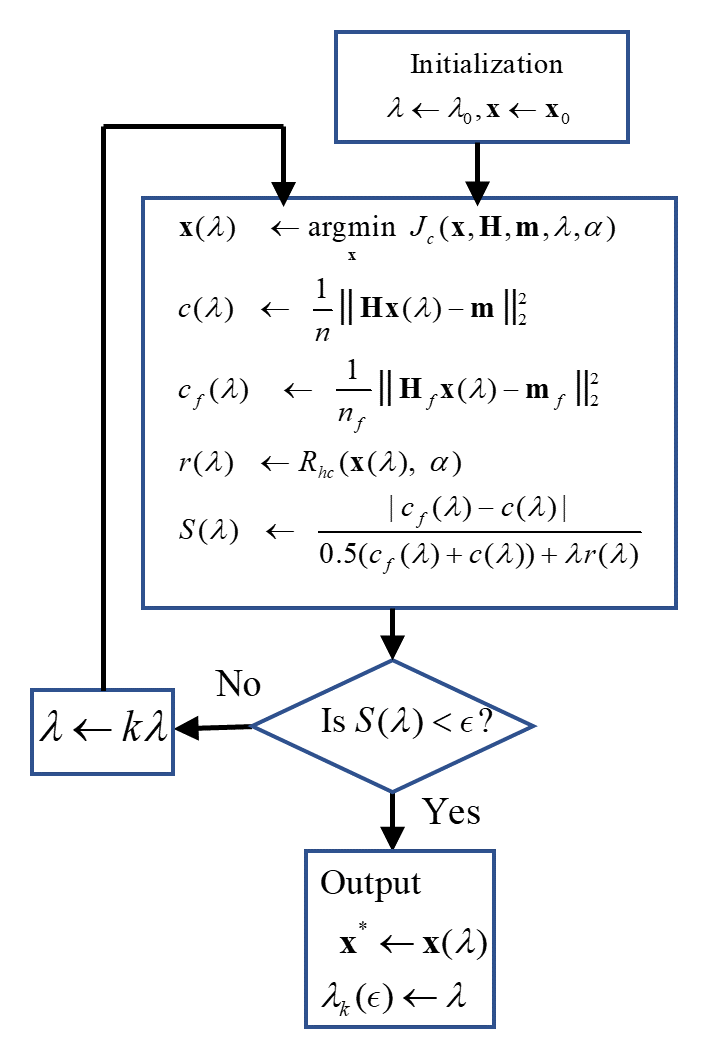}
 \label{fig:scheme1}
 \end{figure}

 Let ${\lambda}(\epsilon)$  be the lowest value of 
 $\lambda$ that satisfy  ${S}(\lambda) \le \epsilon$.  In other words, let 
  ${\lambda}(\epsilon) = min \{\lambda:  {S}(\lambda) \le \epsilon \}$.  Since  $S(\lambda)$
  is monotonic, this is the same as the condition ${S}(\lambda(\epsilon)) = \epsilon$.
  Now, since solving this exactly is impractical,  we use the following strategy:
  we define monotonically increasing  sequence $\{\lambda_i, i =0,1,2,\ldots\}$  such that
  $S(\lambda_0) > \epsilon$,  and find a index $m$ such that $S(\lambda_{m-1}) > \epsilon$  and
  $S(\lambda_{m}) \le \epsilon$. We then declare $\lambda_{m}$ as an approximation for 
  ${\lambda}(\epsilon)$. We define the sequence in the form $\lambda_i = \lambda_0k^i$
  where   $k$ is the real number of the form   $k=1+\delta$ with $\delta$ being a small positive number.
  Here,  since $k$ determines how well $\lambda_m$ approximates  ${\lambda}(\epsilon)$,
  we denotes this approximation by ${\lambda}_k(\epsilon)$. 
  Determining  ${\lambda}_k(\epsilon)$  this way and obtaining the corresponding reconstruction
  can be represented by the schematic given in   Figure \ref{fig:scheme1}.
  In summary,  the algorithm of     Figure \ref{fig:scheme1}   performs reconstruction 
  without the user having to specify the regularization weight directly,  and it jointly determines the
  required regularization weight from the user-specified relative smoothness $\epsilon$. 
  One apparent problem  in the algorithm is that it brings two additional parameters,  viz., 
  $\Delta$ and $k$.  However,  as we will demonstrate in the experiments section,
  these parameters can be set at a nominal value independent of  the measurement vector ${\bf m}$.
 In the schematic of Figure \ref{fig:scheme1},
 we have given computationally efficient formula for  ${S}(\lambda)$  by exploiting the
 fact that the terms $J_c({\textbf{x}}(\lambda), {\bf H}_f, {\bf m}_f, \lambda, \alpha)$  and
 $J_c({\bf x}(\lambda), {\bf H}, {\bf m}, \lambda, \alpha)$ differ only by the data fitting term.
 We also used the fact that ${\cal B}_{[u]}(\cdot)$ will be zero at ${\bf x}(\lambda)$.
 We denote the overall operation of the reconstruction and tracking given in Figure \ref{fig:scheme1}
by $({\bf x}^*,\lambda_k(\epsilon)) \leftarrow 
{\cal T}({\bf x}_0, \lambda_0, k, {\bf H}, {\bf H}_f, {\bf m}, {\bf m}_f, \alpha)$.
 Note that, here,  ${\bf x}_0$ denote the initial candidate for reconstruction and $\lambda_0$ is
 the starting value for used for tacking $\lambda$.

\begin{figure}[htp] 
 	\caption{Regularization weight tracking with partial reconstruction}
 	\centering
 	\includegraphics[width=0.5\textwidth]{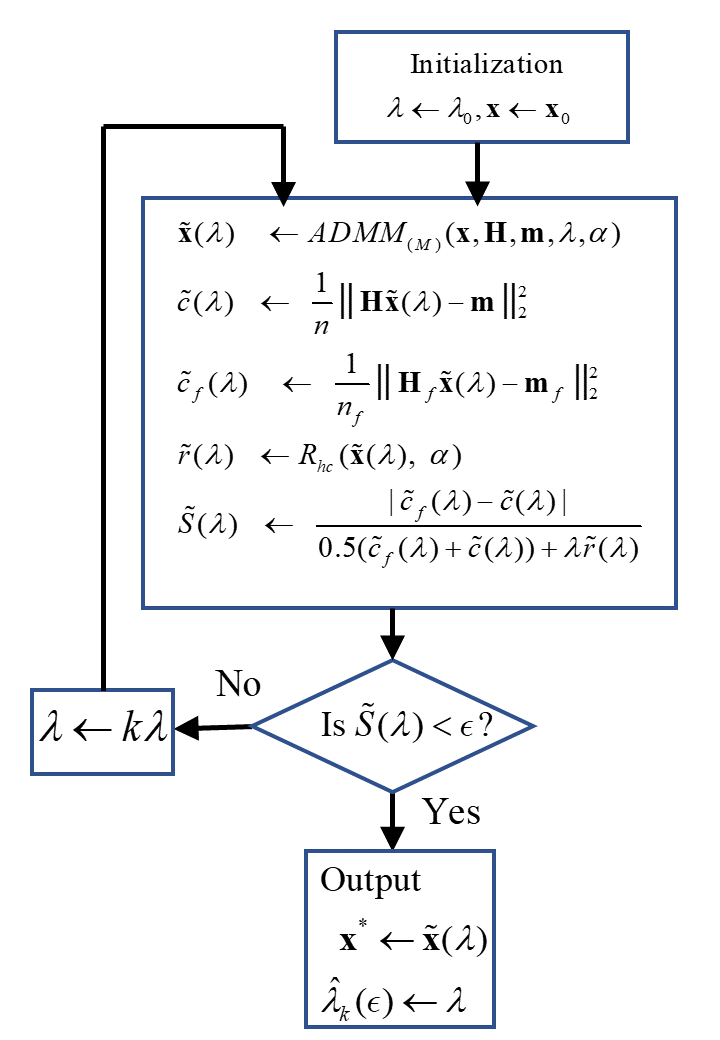}
 	\label{fig:scheme2}
 \end{figure}

 The above scheme only represents an intermediate stage towards developing more pratical
 method, which is our next goal.  The main drawback  in the above scheme is that
 the tracking uses full minimization of the cost for each candidate value of $\lambda$.
 In the remainder of this section,  we develop a more efficient approach that integrates
 the task of determining an approximation for $\lambda_k(\epsilon)$ from a given    $\epsilon$ and
 the task of obtaining the reconstruction.  To this end,  we introduce an approximation for 
 ${S}(\lambda)$, denoted by   $\tilde{S}(\lambda)$,  as given below:
\begin{equation}
\label{eq:sdefapp}
\tilde{S}(\lambda) =  \frac{|\hat{\cal J}_f(\tilde{\textbf{x}}(\lambda), \lambda) - 
\hat{\cal J}(\tilde{\textbf{x}}(\lambda), \lambda)|}
{0.5(\hat{\cal J}_f(\tilde{\textbf{x}}(\lambda), \lambda) +
\hat{\cal J}(\tilde{\textbf{x}}(\lambda), \lambda))},
\end{equation}
where $\tilde{\textbf{x}}(\lambda)$ denotes a partial reconstruction obtained by performing 
a few ADMM iterations, and  $\hat{\cal J}_f(\cdot)$   and $\hat{\cal J}(\cdot)$ denote the costs obtained from 
${\cal J}_f(\cdot)$   and ${\cal J}(\cdot)$    by removing ${\cal B}_{[u]}(\cdot)$.
The main assumption that will lead  the proposed algorithm
is that $\tilde{S}(\lambda)$ will have a similar type of behaviour provided that the number
of ADMM iterations is sufficiently large.  Based on this assumption,  we propose a modification
 to the scheme of Figure \ref{fig:scheme1}, by merging the minimization of $J_c(\cdot)$ with
 tracking for $\lambda$. As first step,  we define the modified tracking denoted by
  $(\hat{\bf x},\hat{\lambda}_k(\epsilon)) \leftarrow 
 {\cal T}_{(M)}({\bf x}_0, \lambda_0, k, {\bf H}, {\bf H}_f, {\bf m}, {\bf m}_f, \alpha, \epsilon)$  that
 is obtained by replacing the minimization ${\bf x}(\lambda) \leftarrow \underset{\bf x}{\operatorname{argmin}} \;
 J_c({\bf x},{\bf H}, {\bf m}, \lambda, \alpha)$  by $M$ cycles of ADMM iterations, 
 denoted by 
 $\tilde{\bf x}(\lambda)  \leftarrow   ADMM_{(M)}({\bf x},{\bf H}, {\bf m}, \lambda, \alpha)$.
 Here ${\bf x}$ in $ADMM_{(M)}({\bf x},{\bf H}, {\bf m}, \lambda, \alpha)$ denotes the starting point
 of the iterations.  
   This scheme is given
 in Figure \ref{fig:scheme2}.    Note that this approximate scheme brings another  parameter, $M$, 
 the number of ADMM iterations performed before each computation of   $\tilde{S}(\lambda)$.
 Again, as we will demonstrate, this can also be fixed at  a nominal value independent of  the measurement vector ${\bf m}$.

\begin{figure}[htp] 
	\caption{Overall proposed algorithm.  ${\cal T}({\bf x}, \lambda, k, {\bf H}, {\bf H}_f, {\bf m}, {\bf m}_f, \alpha, \epsilon)$
	represents the algorithm of Figure \ref{fig:scheme2}}
	\centering
	\includegraphics[width=0.5\textwidth]{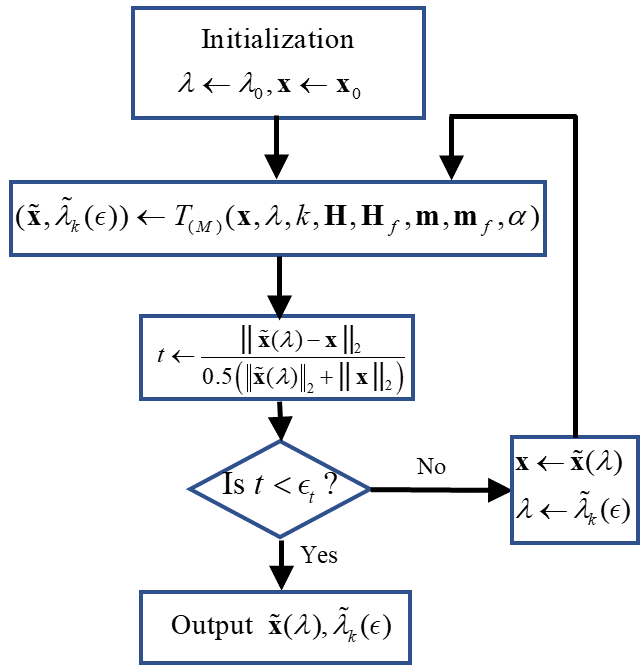}
	\label{fig:scheme3}
\end{figure}

 The final proposed algorithm is the repeated application of this tracking with reconstruction, 
 ${\cal T}_{(M)}(.)$,  where  the pair $(\hat{\bf x},\hat{\lambda}_k(\epsilon))$ returned by each tracking
 becomes the initialisation  to the next tracking.  This process is repeated until 
 relative change in the reconstructed image is lower than the user-defined tolerance $\epsilon_t$. This
 is given in Figure \ref{fig:scheme3}.   Comparing with the original tracking scheme of Figure \ref{fig:scheme1},
 there is no guarantee that this approximate scheme will produce the same results.  However,  as we demonstrate
 in the experiment section,  the result produced by the approximate scheme has a quality that is comparable to that
 of the reconstruction obtained by oracle tuning.    Note that the final parameter, $\epsilon_t$,  is  the termination tolerance.
 As any iterative optimization based image reconstruction, this parameter is set to  small positive real number.

\section{Experimental results}

\begin{figure}[http]
\centering
\includegraphics[width=1\linewidth,trim=4 4 4 4,clip]{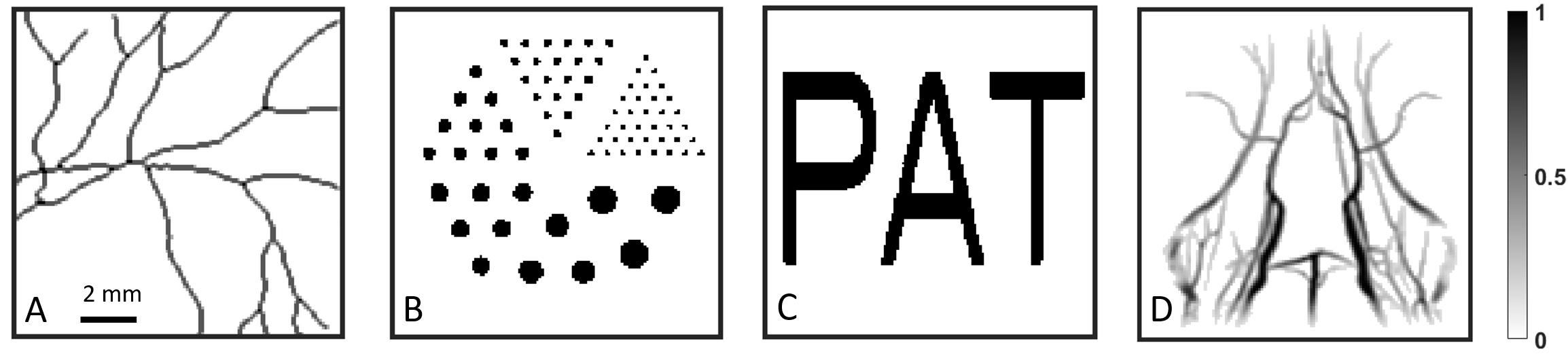}
\caption{Numerical phantoms used for  evaluation of the proposed method}
\label{fig:phantomlist}
\end{figure}

Four numerical phantoms viz. BloodVessel, Derenzo, PAT, and RatBrain,  given in Figure \ref{fig:phantomlist}
were  used to evaluate PAT reconstruction methods.
The first three are synthetic phantoms, whereas the last was derived  from the MRI angiogram of rat brain
published by Pastor et al \cite{pastor2017general}.
All were normalized to the size 128 × 128 with a corresponding physical size of 12.8 mm × 12.8 mm as  models for generating the synthetic data. 
The measurement data was generated using a numerical approximation of forward model described in \cite{liu2016curve} and added with Gaussian noise to form simulated data having SNR levels of 15 dB, 20 dB, 25 dB and 30 dB. 
The number of transducers to mimic the limited data scenario in our experiments was set to  16.  The sound speed in the medium was assumed to be 1.5 mm/$\mu$ s, and we considered the medium to be homogeneous with no dispersion or absorption of sound.  We consider three forms  of reconstruction methods for comparative evaluation:
\begin{itemize}
\item
AR-TR:   Reconstruction using augmented regularization with  proposed  tracking of $\lambda$, i.e.,  the method given
in Figure \ref{fig:scheme3}.
\item
AR-O:  Reconstruction using augmented regularization with   oracle tuning of $\lambda$
\item
TV2-O:  Reconstruction using TV-2  regularization with   oracle tuning of $\lambda$
\end{itemize}
Our primary goal is to demonstrate the power of augmented regularization
with the regularization  weight determined using the proposed   tracking method of Figure \ref{fig:scheme3}.
This is taken care by the above list   in two ways:  (i)  comparing  AR-TR  with AR-O  will show how well the
power of augmented regularization will be accessible in  a practical scenario where the regularization weight
is determined from the measurement vector using the proposed tracking;  it will also quantify the loss incurred 
by  using the approximations involved in  the proposed tacking; 
(ii) comparing AR-TR with TV2-O  will demonstrate that
augmented regularization will still have its advantage  over  the best performing regularization in the literature (TV-2)
despite the above-mentioned loss.

The bound on the relatives smoothness, $\epsilon$,  was fixed at
 $0.06$ uniformly for all measurement sets independent of noise level   and independent of test model used for
simulating measurements.  Further, the other parameter involved in the tracking algorithm were set 
as follows: $\Delta=0.1$,  $k=1.05$, and $M=50$. Again these settings were kept fixed for all input data sets.
The relative weight $\alpha$   was kept fixed at $0.5$ uniformly for all data sets as we did in our original work
\cite{rejesh2020photo}.  
Finally, the termination tolerance  $\epsilon_t$  was set to $10^{-4}$.    The results are  presented in  Table 1.
We observe the following from the tables: 
\begin{itemize}
\item
SSIM scores of reconstructions obtained from AR-TR  are comparable to
that of the reconstructions obtained from AR-O; in most of cases, the difference
is in the third decimal point,  and in rare cases, the difference
is in the second decimal point.  This is significant because AR-O uses the model to determine the value for $\lambda$ whereas
AR-TR does not. 
\item
  SSIM scores of reconstructions obtained from AR-TR are significantly
better than that of TV2-O,  although AR-TR does not use the test model for determining the value for $\lambda$ as opposed to
TV2-O, which uses the model to determine the value for $\lambda$.
\end{itemize}
By noting the fact that  the required $\lambda$ values for AR-TR were tracked from  a fixed value of $\epsilon$, 
we conclude  from  the first observation that  the proposed
 method provide a practical solution to the problem of determining the regularization weight for PAT image reconstruction. 
 The second observation imply that the   advantage of augmented regularization over the standard TV regularization
 is accessible via a practical method for tuning $\lambda$.  Note that in our previous work where we introduced the 
 augmented regularization,  we demonstrated its superiority over the standard TV only by using oracle tuning.

 Note that, among the   published methods for regularization weight tuning
with total variation regularization,  the method of Xue et al. \cite{fengxue}  is closest competitor for the
the proposed  method, AR-TR.  However,   comparison with TV2-O  eliminates the need to compare with
method of  Xue et al.   because TV2-O  is bound to  give reconstruction that is better than that of
the method of Xue et al.   due to  the oracle tuning  in TV2-O.   Besides,  the method of  Xue et al.
is based on Monte Carlo simulations,  and implementing it for PAT reconstruction will be impractical
as PAT forward model   is more complex than the  blurring model considered by  Xue et al.
 In the remainder of this section,  we will display images corresponding to two selected cases from Table 1 and Table 2.

\begin{table}[ht]
\centering
\begin{tabular}{ |c|c| c| c|c|   }
\hline%\multirow {2}{*} { SNR }  & \multicolumn{2}{|c|}{16 trans.}  & \multicolumn{2}{|c|}{16 trans.} & \multicolumn{2}{|c|}{16 trans.}  \\\cline{2-7}
  Phtm. & SNR (dB) & AR-TR  & TV2-O & AR-O    \\
 \hline	
 \multirow{4}{*}{\rotatebox{90}{RatBr}} & 15 &  .916 & .791  &  .919                  \\
 & 20 & .955	& .877	& .957		 \\
& 25 & .977	& .925	& .979	 	 \\
 & 30 & .986	& .956	& .988		\\  \hline
 \multirow{4}{*}{\rotatebox{90}{BldVes}} &  15 & .919         &  .719     & .928  \\   
 & 20 &  .952	         & .827	& .964 \\  
 & 25 &  .980	        & .881	& .983\\  
 & 30 &  .993	& .945	& .993\\   \hline
  \multirow{4}{*}{\rotatebox{90}{Derz}} & 15 &  .995 & .959 & .997    \\   
 & 20 &  .999 & .977 & .999   \\   
 &  25  & .999 & .986 & .999   \\   
  & 30  &  .999 & .989 & .999   \\  \hline
   \multirow{4}{*}{\rotatebox{90}{PAT}} & 15 & .985 & .946 & .985 \\   
 & 20 & .990 & .962 & .990 \\    
 & 25 & .990 & .973 & .992 \\     
 & 30 & .991 & .978 & .994  \\   \hline
 \end{tabular}
 \caption{ SSIM scores of reconstruction with varying input noise levels with 16 transducers and relative
  smoothness parameter ($\epsilon$) value of 0.06.
 }
 \label{table:1}
\end{table}

%\begin{figure}[http]
%\centering
%\includegraphics[width= 0.5\linewidth]{Fig_2.png}
%\caption{a) Schematic diagram of PA data acquisition geometry with ultrasound transducers 
%(shown by dots) around the imaging region of 12.8 mm $\times$ 12.8 mm.} %The computational imaging grid size is   51.2 mm $\times$  51.2 mm.}
%\label{fig:synthscheme}
%\end{figure}

\begin{figure}[H]
 \centering
 \includegraphics[width=0.7\linewidth]{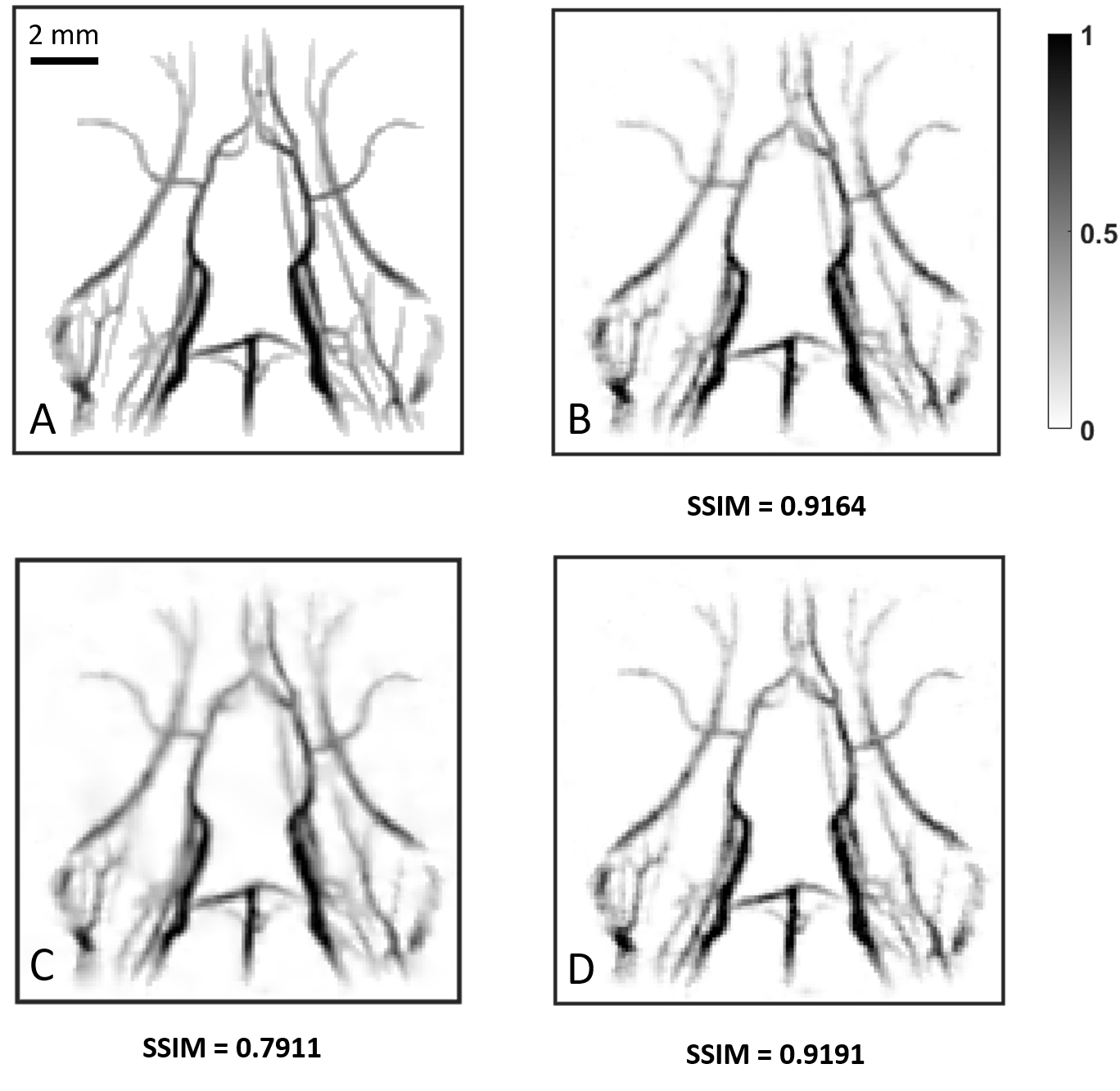}
 \caption{Comparison of reconstructions obtained from simulated data with 16 transducers and 15 dB measurement noise. (A): RatBrain phantom model (the maximum initial Pressure rise 
 is assumed to be 1Pa);  
(B):  Reconstruction obtained by AR-TR;
 (C) Reconstruction obtained TV2-O;
  (D): Reconstruction obtained by AR-O.
  }
  \label{fig:reccomp1}
\end{figure}
\begin{figure}[ht]
 \centering
 \includegraphics[width=0.7\linewidth]{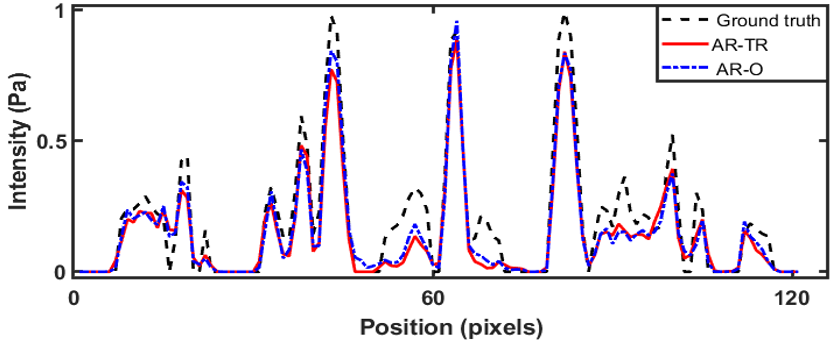}
 \caption{Scan line based intensity profiles of reconstructed images from Fig:\ref{fig:reccomp1}.  
  }
  \label{fig:fig3scan}
\end{figure}

Figure   \ref{fig:reccomp1}  compares the reconstructed results for the RatBrain phantom with the measurement set corresponding
to SNR level of 15 dB. The   original  image  is given in Figure  \ref{fig:reccomp1}.A while Figure  \ref{fig:reccomp1}.B,  
\ref{fig:reccomp1}.C and  \ref{fig:reccomp1}.D gives the reconstructed images from AR-TR,  TV2-O, and AR-O  respectively. 
Clearly, AR-TR  outperforms TV2-O, which is also confirmed by the fact that  the SSIM score  of  AR-TR's reconstruction
is 0.125  higher than that of TV2-O's reconstruction.    Further,    AR-TR's  reconstruction has almost the same quality as
that of   AR-O.     This is confirmed by the  fact that the SSIM score of AR-O's reconstruction is only 0.003
higher than that of AR-TR's reconstruction.    Figure \ref{fig:fig3scan}  shows the scan-line based intensity profile of the reconstructions 
of   Figure  \ref{fig:reccomp1} . The intensity profiles show that  the proposed method, AR-TR,  follows the ground truth as good as
 the oracle tuning method,  AR-O,   which is also confirmed by the small difference in the SSIM scores.

\begin{figure}[H]
 \centering
 \includegraphics[width=0.7\linewidth]{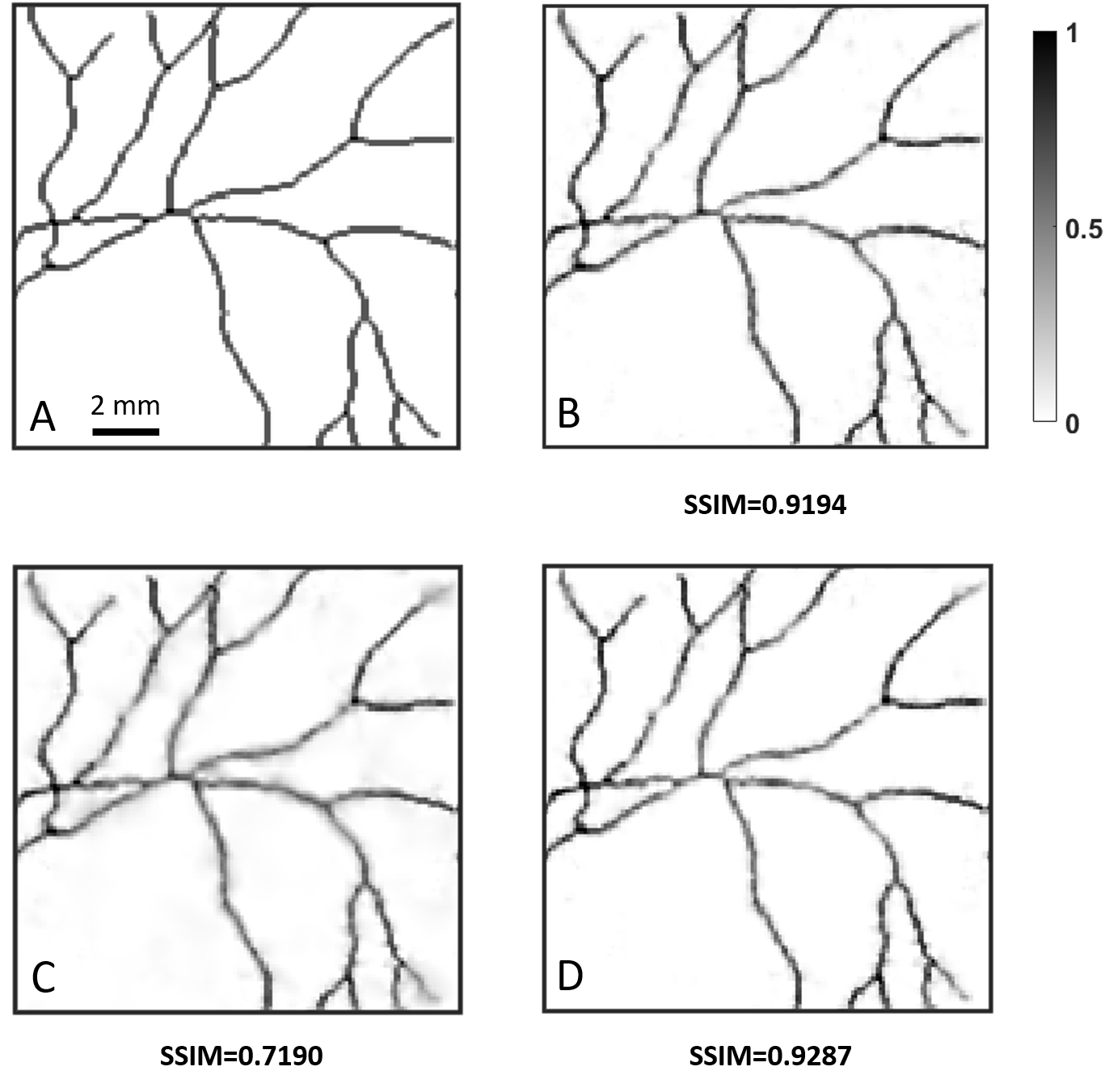}
 \caption{Comparison of reconstructions obtained from simulated data with 16 transducers and 15 dB 
 measurement noise. (A):BloodVessel phantom model (the maximum initial Pressure rise 
 is assumed to be 1Pa);  
(B):  Reconstruction obtained by AR-TR;
 (C) Reconstruction obtained TV2-O;
  (D): Reconstruction obtained by AR-O.  }
  \label{fig:reccomp2}
\end{figure}

\begin{figure}[ht]
 \centering
 \includegraphics[width=0.7\linewidth]{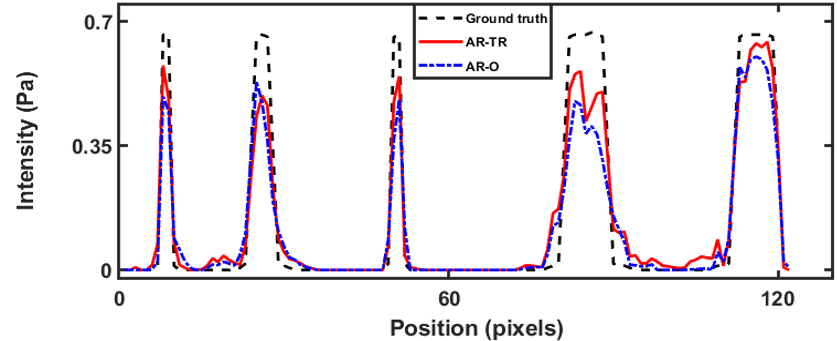}
 \caption{Scan line based intensity profiles of reconstructed images from Fig:\ref{fig:reccomp2}.  
  }
  \label{fig:fig4scan}
\end{figure}

Next, for another closer view,  we select the reconstruction corresponding to  measured data with 15 dB  SNR obtained using BloodVessel phantom.  The images are given in  Figure \ref{fig:reccomp2}. Figure \ref{fig:reccomp2}.A shows the   BloodVessel phantom and Figures 
\ref{fig:reccomp2}.B, \ref{fig:reccomp2}.C and \ref{fig:reccomp2}.D gives the reconstructions from the AR-TR, TV2-O,  and AR-O 
respectively. Here also,  AR-TR is able to produce a reconstruction with a quality that is comparable to the quality of the reconstructions produced by AR-O, which is confirmed by a mere difference of 0.009 in SSIM.  Further,   AR-TR retains the advantage that augmented regularization has over the TV-2 regularization,  which is confirmed by the improvement of 0.2 in SSIM score. Figure \ref{fig:fig4scan} shows the scan-line based intensity profile of the reconstructions in Figure \ref{fig:reccomp2}. Here also, our method closely follows the ground truth as good as the oracle tuning method, which is also confirmed by the small difference in the SSIM scores.

\section{Conclusion}

Currently, there is no method for determining the required regularization weight  from measured data
 for PAT image reconstruction with sparsity  regularization;  users  of a  sparsity  method have to determine
  the weight by trail and  error approach
where the reconstruction is performed for a series of values for the regularization weight, $\lambda$,
 with a visual feed back on the quality of the reconstructed image.  This  is impractical  because  the appropriate value of 
 $\lambda$ vary drastically  with variations in the noise level and the structural contents of the underlying image. 
 In this paper, we addressed this problem  by introducing a novel relative smoothness parameter  as an implicit 
 function of $\lambda$  that  is  defined through a partial  reconstruction.   We constructed  an iterative method
 that jointly determines, for given value of relative smoothness,  the  corresponding  $\lambda$   and the required
 reconstruction with augmented sparsity regularization.   We demonstrated experimentally that we can get good quality reconstructions from this new method
 by setting the relativeness parameter at a fixed nominal value independent of the noise level and the nature of  the
 underlying image. We showed that the SSIM scores of  reconstructions obtain this way were close to the SSIM
 scores of reconstructions obtained   from the scheme where the  underlying images  themselves  were used for  determining
 the correct value for $\lambda$.   Further, other parameters involved in the definition of the relative
 smoothness measured also have been set at fixed values independent of noise level and the underlying image.
 This means that, in practical point of view, our method solves the
 problem of determining the regularization weight from measured data.
\bibliography{pattune_jinst_v4}
%\bibliography{pattune_jinst_v3}
\bibliographystyle{iopart-num}

\end{document}